\documentclass[manuscript]{aastex}

\usepackage{color}
\usepackage{amsmath}

\shorttitle{}
\shortauthors{Monje et al.}

\begin{document}


\title{Hydrogen chloride in diffuse interstellar clouds along the line of sight to W31C (G10.6-0.4)}


\author{R. R. Monje, D. C. Lis} 
\affil{California Institute of Technology, MC 301--17, 1200 E. California Blvd., Pasadena, CA  91125-4700, USA}
\email{raquel@caltech.edu}
\author{E. Roueff}
\affil{Observatoire de Paris-Meudon, LUTH UMR 8102, 5 Pl. Jules Janssen, F-92195 Meudon Cedex, France}
\author{M. Gerin, M. De Luca}
\affil{LERMA, CNRS, Observatoire de Paris and ENS, France}
\author{D. A. Neufeld}
\affil{Department of Physics and Astronomy, Johns Hopkins University, 3400 North Charles Street, Baltimore, MD 21218, USA}
\author{B. Godard}
\affil{Departamento de Astrof\'{\i}sica, Centro de Astrobiolog\'{\i}a (CAB), INTA-CSIC, Crta. Torrej\'on km 4, 28850, Torrej\'on de Ardoz, Madrid, Spain}
\author{ and T. G. Phillips} 
\affil{California Institute of Technology, MC 301--17, 1200 E. California Blvd., Pasadena, CA  91125-4700, USA}




\begin{abstract}

We report the detection of hydrogen chloride, HCl, in diffuse molecular clouds on the line of sight towards the star-forming region W31C (G10.6-0.4). The $J=1-0$ lines of the two stable HCl isotopologues, H$^{35}$Cl and H$^{37}$Cl, are observed using the 1b receiver of the Heterodyne Instrument for the Far-Infrared (HIFI) aboard the Herschel Space Observatory. The HCl line is detected in absorption, over a wide range of velocities associated with diffuse clouds along the line of sight to W31C. The analysis of the absorption strength yields a total HCl column density of few 10$^{13}$~cm$^{-2}$, implying that HCl accounts for $\sim 0.6$~\% of the total gas-phase chlorine, which exceeds by a factor of $\sim$~6 the theoretical model predictions. This result is comparable to those obtained from the chemically-related species H$_2$Cl$^+$ and HCl$^+$, for which large column densities have also been reported on the same line of sight.  The source of discrepancy between models and observations is still unknown; however, the detection of these Cl-bearing molecules, provides key constraints for the chlorine chemistry in the diffuse gas.

\end{abstract}


\section{Introduction}

Hydride molecules play an important role in interstellar chemistry, as they are often stable end points of chemical reactions, or represent important intermediate stages of the reaction chains theorized to form gas-phase molecules. This makes hydrides a sensitive test of these chemical models, as well as potential tracers of other molecules of interest e.g. molecular hydrogen. Due to their small moment of inertia, hydrides have their fundamental rotational lines in the submillimeter band. The Heterodyne Instrument for the Far-Infrared (HIFI) on board the Herschel Space Observatory is providing invaluable data on interstellar chemistry in general, and in particular on hydride molecules within the Milky Way and local galaxies. Halogen atoms fluorine (F) and chlorine (Cl), with estimated solar abundances of 3.6 $\times$ 10$^{-8}$ and 3.2 $\times$ 10$^{-7}$ relative to hydrogen (Asplund et al. 2009), are of special interest because they are the major atoms (neutral F and ionic Cl$^+$) in diffuse environments that react exothermically with molecular hydrogen to form hydride molecules (HF and HCl). HF and HCl are strong bound systems, only destroyed by photodissociation, reactions with He$^{+}$, H$^{+}_{3}$, and C$^{+}$ and, in the case of HCl, also by photoionization.
Chlorine chemistry has been determined by extensive theoretical and observational work \citep{Jur74,Dal74,van86,Bla86,Sch95,Fed95}. In diffuse clouds, Cl atoms can be ionized by UV photons at wavelengths between 91.2 and 95.6 nm \citep{Jur74}. The resulting ion, Cl$^{+}$, reacts rapidly exothermically (by 0.17 eV) with H$_2$ to form HCl$^{+}$ 
\begin{equation}
\centering
	\rm Cl^{+}~+~H_2~\rightarrow~HCl^{+}~+~H,
\end{equation}
which in turn reacts with H$_2$ to form chloronium, H$_2$Cl$^+$
\begin{equation}
\centering
	\rm HCl^{+}~+~H_2~\rightarrow~H_2Cl^{+}~+~H.
\end{equation}
HCl is then formed through dissociative recombination (DR) of H$_2$Cl$^+$
\begin{equation}
\rm  H_2Cl^{+}~+~e~\rightarrow \left\{ 
 \begin{array}{l}
\rm	Cl + H_2\\
\rm	HCl + H.\\
 \end{array} 
 \right.
\end{equation}

From reactions (1)~--~(3), HCl is expected to be abundant in regions containing both H$_2$ and chlorine-ionizing photons. 

Observations of HCl have been limited to ground-based observatories at high-altitude sites, under extreme good weather conditions, or space-missions (due to its large rotational constant which places the ground state transition, at 625.9187 GHz, near a strong atmospheric water absorption line). The first detections of interstellar chlorine-containing molecules were obtained using the NASA's Kuiper Airborne Observatory (KAO). \cite{Bla85} observed the HCl ground-state $J=1-0$ rotational transition in emission in OMC-1 with angular resolution of about 2$\arcmin$. \cite{Zmu95} detected the line in absorption towards Sagittarius B2. Following these results, ground-based observations have lead to the detection of the H$^{35}$Cl and H$^{37}$Cl isotopologues, in environments such as evolved stars and active star-forming regions \citep{Sch95,Sal96,Pen10}. 

Cl chemistry in diffuse clouds on the other hand, has not been fully constrained because of the paucity of observations of Cl-bearing species in the interstellar medium. Prior to Herschel/HIFI, HCl in diffuse clouds was observed mainly by means of ultraviolet absorption studies leading only to a tentative detection towards $\zeta$ Oph \citep{Fed95}, with an HCl column density of $2.7 \times 10^{11}$~cm$^{-2}$. Given an atomic chlorine column density of 3.0 $\times$ 10$^{14}$ ~cm$^{-2}$ for this line of sight \citep{Fed95}, the $N$(HCl)/$N$(Cl) ratio is $9 \times 10^{-4}$. This result is in good agreement with the theoretical models of the Cl chemistry (\citealt{van86} and more recently \citealt{Neu09}, hereafter NW09). 
The NW09 models predict column densities in the range of 10$^{10}$~--~10$^{12}$ cm$^{-2}$ for HCl, H$_2$Cl$^+$ and HCl$^+$, for a diffuse molecular cloud of density $n_{\rm H} = 10^{2.5}$ cm$^{-3}$ and $\chi$$\rm_{UV}$ in the range of 1~--~10, where $\chi$$\rm_{UV}$ is the UV radiation field normalized with respect to the mean interstellar value (\citealt{Dra78}; see Figures 6 and 7 in NW09). Model predictions also identified H$_2$Cl$^+$ and HCl$^{+}$ as potentially detectable Cl-bearing species. 

The Herschel/HIFI instrument has indeed allowed observations for the first time of these two new Cl-bearing species in the interstellar medium, as well as the detection of HCl in diffuse clouds \citep{Lis10,deL12} and in protostellar shocks \citep{Cod12}. The molecular ions H$_2$Cl$^+$ and HCl$^+$ have been detected in diffuse clouds on the lines of sight to bright submillimeter continuum sources. H$_2$Cl$^+$ has been detected in absorption towards NGC 6334I, Sagittarius B2 (S) \citep{Lis10}, W31C, Sgr A and in emission towards Orion Bar, Orion South \citep{Neu12}. The H$_2$Cl$^+$ column densities obtained from these studies, imply that chloronium accounts for $\sim$ 4~--~12~\% of chlorine nuclei in the gas phase. This result is at least a factor $\sim$ 10 larger than that predicted by the chemical models, which predict a H$_2$Cl$^+$ and HCl$^+$ joint contribution of $\sim$ 2~\% to the total gas-phase chlorine budget \citep{Neu12}. Similar results have been derived from observations of HCl$^+$ towards W31C and W49N \citep{deL12}, where HCl$^+$ has been detected in absorption with large column densities suggesting a 3~--~5~\% contribution of this species to the total gas-phase chlorine content. 

The discrepancy between the chlorine chemistry models and observations is very puzzling; diffuse clouds contain only simple molecules, and the number of reactions involved in describing their abundances is fewer than in dense molecular clouds. Therefore, a test of the basic interstellar chemistry should be possible, establishing much higher standards for the modeling of diffuse clouds compared to those of dense clouds. The diffuse cloud models thus should be able to reproduce the measured abundances to a factor of two or better \citep{van90}.  

In this paper, we report the result of Herschel/HIFI observations of HCl $J=1-0$ on the line of sight towards W31C (G10.6-0.4). These observations were previously attempted by \cite{Pen10} using the Caltech Submillimeter Observatory (CSO), but yielded no HCl detection due to the high rms of $\sim$ 0.15 K at a velocity resolution of 0.31 km s$^{-1}$ in their observations. W31C is an extremely active region of high-mass star formation, and one of the three bright HII regions of the W31 complex, harboring an extremely luminous submillimeter and infrared continuum source \citep[L$\rm_{IR}$ $\sim$ 10$^7$ L$_\odot$,][]{Wri77}. With a kinematic distance of 4.8$^{+0.4}_{-0.8}$ kpc \citep{Fis03}, the W31C line of sight intersects several foreground molecular clouds from the Milky Way spiral arms. The source location and its strong continuum flux, make W31C one of the best sources towards which to carry out absorption line studies. We compare the Herschel/HIFI results with current chlorine chemistry models and with observations of other Cl-bearing molecules detected along the same line of sight and toward the galactic center source Sgr B2(S). 

\section{Observations}

The observations were performed with the Herschel/HIFI instrument \citep{deG10} on March 2 2010, as part of the PRISMAS (Probing Interstellar Molecules with Absorption lines Studies) guaranteed time key program. Both H$^{35}$Cl and H$^{37}$Cl $J=1-0$ lines, with rest frequencies of  625.918 and 624.977 GHz respectively, were observed simultaneously using the HIFI band 1b receiver, with 3 shifted local oscillator (LO) settings and the wide band spectrometer (WBS). The WBS provides a spectral resolution of 1.1 MHz 
over a 4~GHz Intermediate Frequency (IF) bandwidth. The dual beam switch (DBS) observing mode was used with reference beams located 3$\arcmin$ on either side of the source position ($\alpha_{J2000}$ = 18$^{h}$10$^{m}$28.700$^{s}$ and $\delta_{J2000}$ = -19$^{\circ}$55$\arcmin$50.00$\arcsec$) along an East-West axis. 


The data have been reduced using the Herschel Interactive Processing Environment (HIPE) \citep{Ott10} with pipeline version 5.2. The resulting Level 2 double sideband spectra were exported into FITS format for subsequent data reduction and analysis using the IRAM GILDAS package\footnote{http://iram.fr/IRAMFR/GILDAS/}. The spectra obtained at different LO setting and for each polarization were inspected for possible contamination by emission lines from the image sideband and averaged to produce the final spectra shown in Figure \ref{fig1}. Beam measurements, reported on November 17 of 2010, towards Mars at 610 GHz give a main beam ($\eta_{mb}$) of 0.744 and 0.764, for the Horizontal (H) and vertical (V) polarization, respectively. The full width at half maximum (FWHM) HIFI beam size at the HCl $J=1-0$ frequency is $\sim$~33\arcsec.   


\section{Results}
\begin{figure}
\centering
\includegraphics[angle=0,scale=.60]{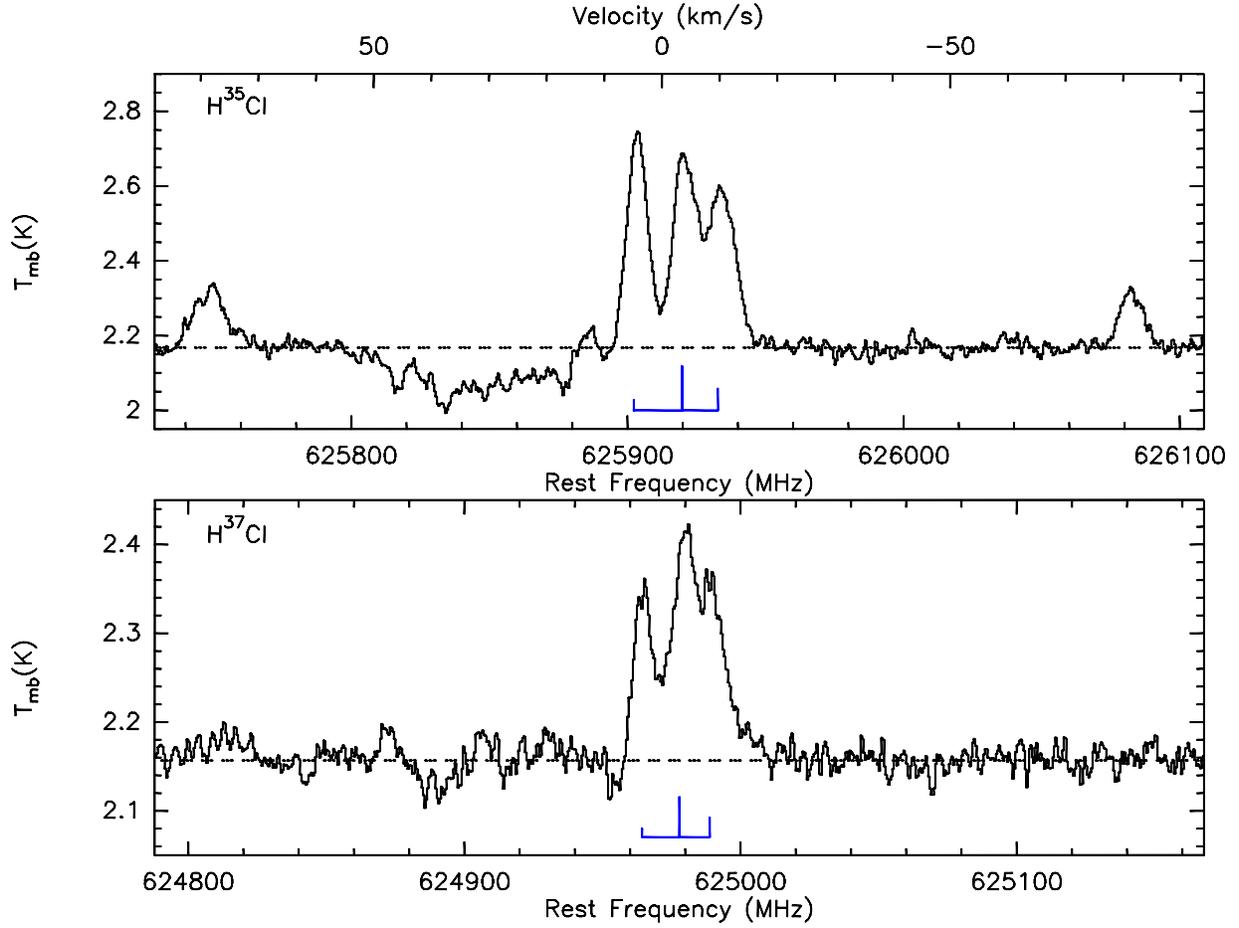}
\caption{Spectra of the $J=1-0$ transition of H$^{35}$Cl (\textit{upper panel}) and H$^{37}$Cl (\textit{lower panel})  towards W31C. The velocity scale in the upper axis is with respect to the frequency of the main ($F~=~5/2~-~3/2$) hyperfine component of both isotopologues. \label{fig1}The blue lines indicate the position of the HFS components. }
\end{figure}

The spectra obtained for the two stable hydrogen chloride isotopologues, H$^{35}$Cl and H$^{37}$Cl towards W31C are shown in Figure \ref{fig1}, upper and lower panels respectively. The data quality is excellent with a double sideband continuum antenna temperature of $\sim$~1.6~K, and a rms noise of $\sim$~6~mK. Figure \ref{fig1} shows the H$^{35}$Cl and H$^{37}$Cl emission lines from the background source and a clear absorption feature in the H$^{35}$Cl spectrum, at velocities $\sim$~10~--~50~km~s$^{-1}$. The emission lines from the source will be discussed in an extended paper, which will include the HCl observations towards all PRISMAS sources (Monje et al. in prep.). The H$^{35}$Cl absorption feature is analogous to the one seen in the spectra of other molecules such as HF, CH and H$_2$O \citep[see Figure 2 in][]{Neu12} and corresponds to the foreground clouds on the line of sight towards W31C. The foreground clouds are known to harbor low density and cool molecular gas (T$\rm_{kin}$ $\sim$ 50~-~70~K from H$_2$ and $\sim$~100~K from H I), where molecular excitation can be highly subthermal, with excitation temperatures close to the cosmic microwave background radiation $\sim$~2.73~K \citep{God10}. Consequently, molecular emission lines are very weak and these clouds are best studied through absorption spectroscopy. 

\begin{figure}
\centering
\includegraphics[angle=0,scale=.60]{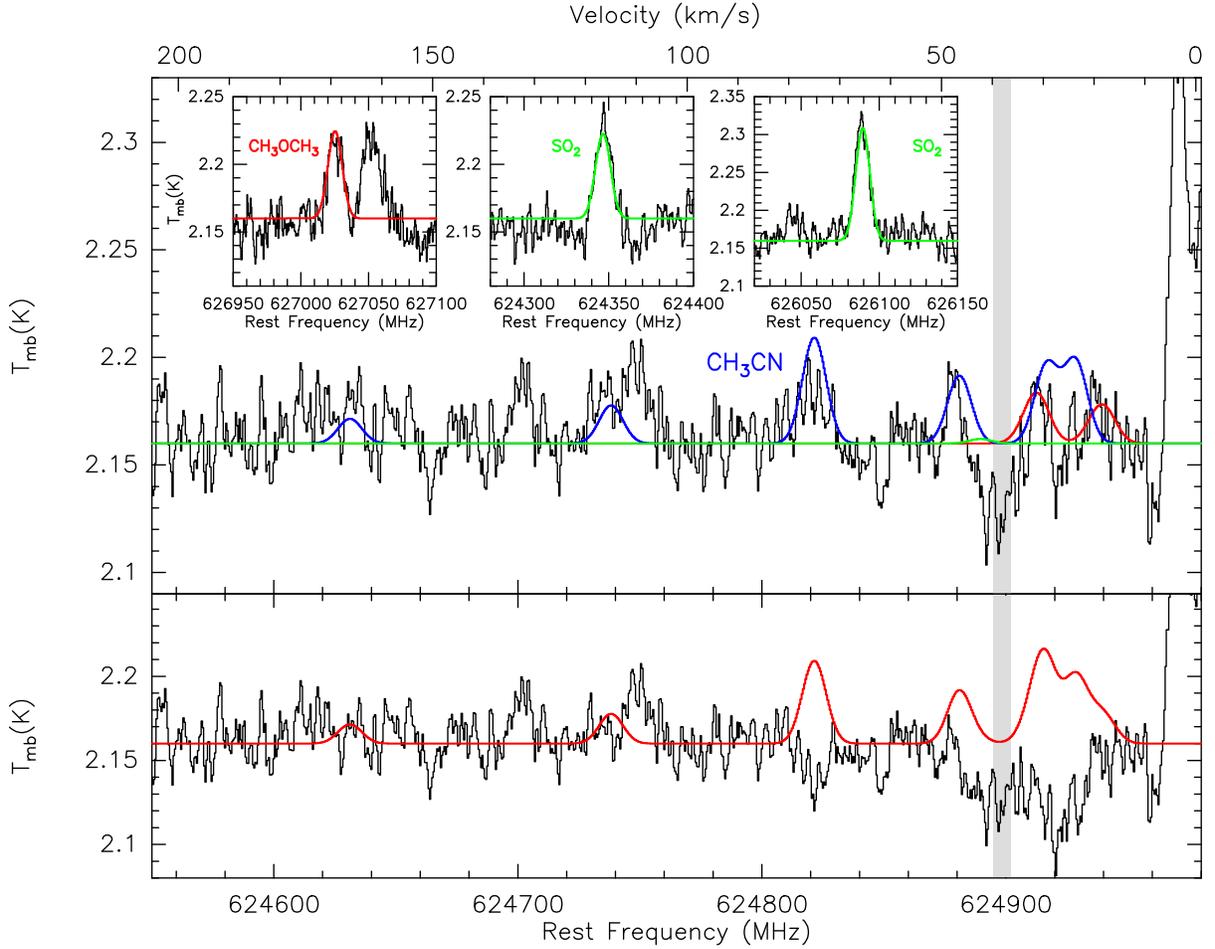}
\caption{Upper panel, XCLASS LTE models of the dimethyl ether (red), methyl cyanide (blue) and sulfur dioxide (green) emission lines blended into the H$^{37}$Cl absorption line spectrum, using the parameters given in Table 1. Additional lines (within the observations bandwidth) used in the model are shown in small windows at the top of the plot. The horizontal axis is in frequency units, with the corresponding velocity scale plotted on the top horizontal axis for consistency with Figure \ref{fig1}. Lower panel, for illustration purposes only shows the resulting H$^{37}$Cl spectrum (black) after subtracting the contamination from emission lines shown in red. The grey area highlights the emission line contamination-free area within the H$^{37}$Cl absorption line spectrum.\label{fig2}}
\end{figure}

\begin{figure}
\centering
\includegraphics[angle=-90,scale=.60]{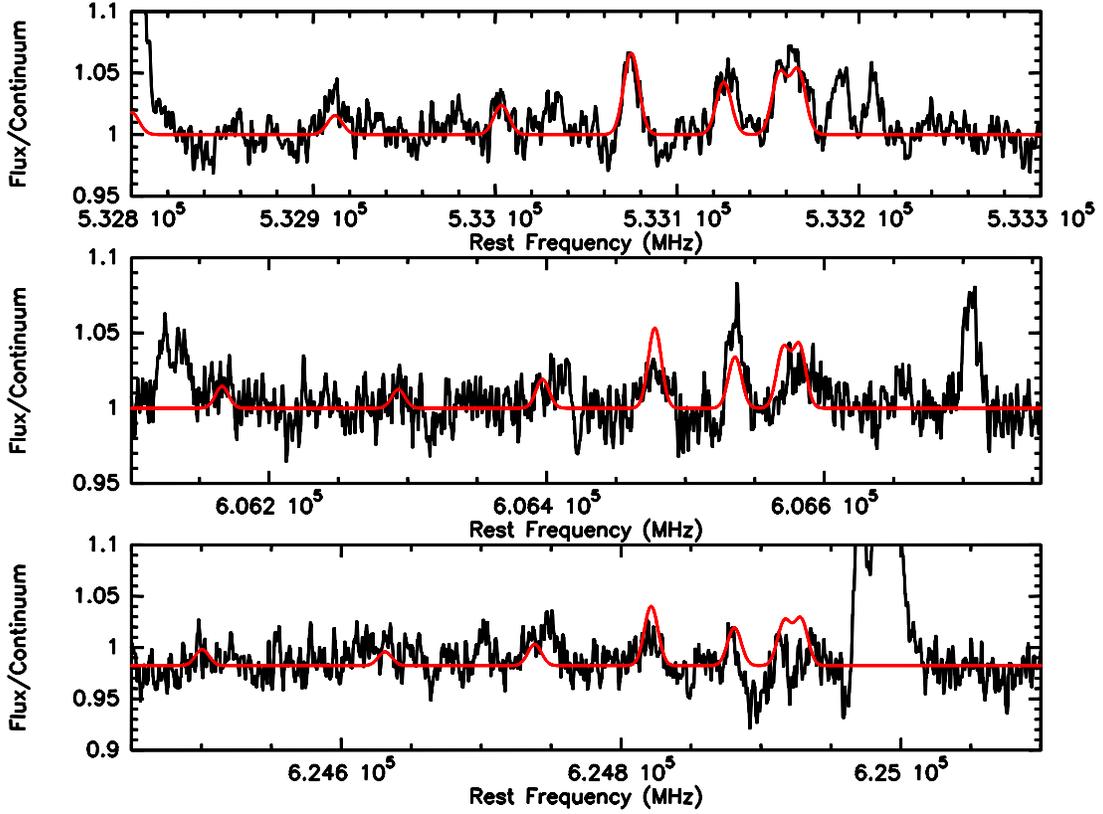}
\caption{LTE model of CH$_3$CN (red), using CH$_3$CN transition lines $J=29-28$ (upper panel) at 533.123 GHz,  $J=33-32$ (middle panel) at 606.533 GHz and $J=34-33$ (lower panel) at 624.878 GHz. The fit parameters are shown in Table 1. \label{fig3}}
\end{figure}

The corresponding H$^{37}$Cl absorption line is contaminated by interfering emission of dimethyl ether (CH$_3$OCH$_3$), methyl cyanide (CH$_3$CN, v=0) and sulphur dioxide (SO$_2$). The emission lines are modeled with the program XCLASS \citep{Sch98,Com05}, which assumes local thermodynamic equilibrium (LTE) and uses the rest frequencies provided by the Cologne Database for Molecular Spectroscopy (CDMS) \citep{Mul01,Mul05} and the Jet Propulsion Laboratory (JPL) molecular spectroscopy database \citep{Pic98}. The result from the line fits exhibits two relatively strong CH$_3$CN and CH$_3$OCH$_3$ lines and a weak SO$_2$ line within the H$^{37}$Cl absorption spectrum, see Figure \ref{fig2}. To obtain the best LTE fit for CH$_3$OCH$_3$ and  SO$_2$ additional lines within the same LO tuning are used, see Figure \ref{fig2}. For the CH$_3$CN additional transition lines at 533.123 GHz ($J=29-28$) and 606.533 GHz ($J=33-32$) are used, see Figure \ref{fig3}. The parameters used in the LTE model, shown in Table 1, assume that the CH$_3$CN and SO$_2$ are tracers of the hot core \citep{Bel11} while the CH$_3$OCH$_3$ arises from the more extended envelope. However, the W31C region is known to be a complex source with outflow activity from embedded high-mass protostars, methanol masers, a dense rotating toroid and infalling material from the molecular envelope. The LTE emission line fit model is thus an approximation and tighter constraints using the combination of key molecules such as methanol, formaldehyde, CO, CS and SO and more complex chemical modeling are needed. Due to the uncertainties of the modeling, we only use the contamination-free spectra (grey shaded area in Figure \ref{fig2}) to derive an estimate of the $^{35}$Cl/$^{37}$Cl isotopic ratio, as described below. 

\begin{figure}
\centering
\includegraphics[angle=0,scale=.60]{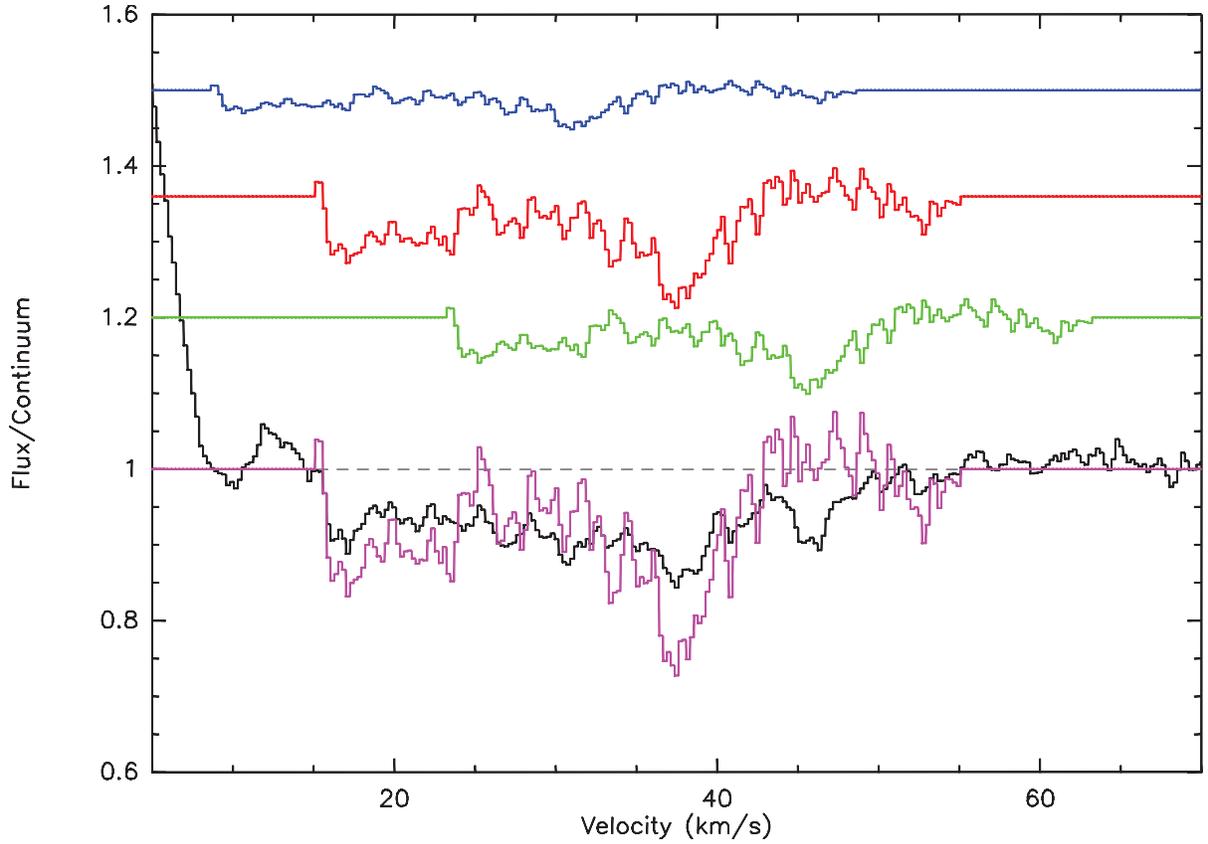}
\caption{Spectra associated with each HFS components of the H$^{35}$Cl line: $F_1=3/2-3/2$, $F=2-2$ (\textit{green}), $F_1=5/2-3/2$, $F=3-2$ (\textit{blue}); $F_1=1/2-3/2$, $F=1-2$ (\textsl{red}) (vertical offset introduced for clarity),  the sum of the three components (\textit{magenta}) and the observed spectrum (\textit{black})  \label{fig4}}
\end{figure}
\begin{figure}
\centering
\includegraphics[angle=0,scale=.60]{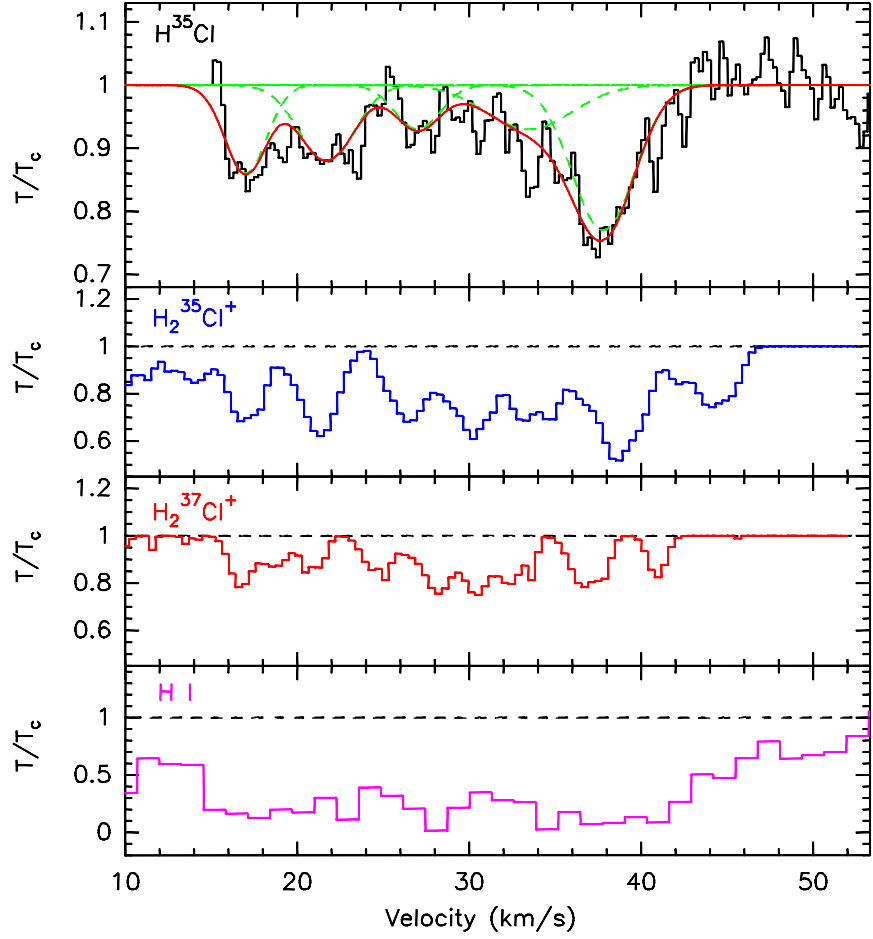}
\caption{Top panel, multiple Gaussian fit (red) to the hyperfine deconvolved spectrum of H$^{35}$Cl $J=0-1$ (black) associated with foreground clouds on the line of sight towards W31C. For comparison, we show the HFS deconvolved spectra of H$_2$$^{35}$Cl$^+$ (blue), H$_2$$^{37}$Cl$^+$ (red) from \cite{Neu12} and H~I (magenta) from \cite{Fis03}.  \label{fig5}}
\end{figure}

The rotational levels of HCl have a hyperfine splitting, caused by the interaction of the quadrupole moment of the chlorine nucleus (I~=~3/2) and the electric field. The $J = 0 - 1$ transition is thus split into three components with relative strengths of 50~\%, 16.7~\%, and 33.3~\%, and with velocity offsets ($\Delta$v$\rm_h$) of 0, -6.3, and +8.2~km~s$^{-1}$ for H$^{35}$Cl, and 0, -5 and 6.5 km s$^{-1}$ for H$^{37}$Cl. The $\Delta$v$\rm_h$ is larger than the velocity difference between consecutive velocity components, producing overlaps between different velocity components that results in a complex profile. In order to determine the velocity structure of the foreground absorbing gas, we extract the signal associated with each HFS component using the numerical procedure described in \cite{God12}.
In Figure \ref{fig4}, we present the H$^{35}$Cl spectrum, with flux normalized with respect to the single sideband continuum (2T$\rm_L$/T$\rm_C$~-~1, where T$\rm_{L}$/T$\rm_{C}$ is the line-to-continuum ratio) in the velocity range from $\upsilon$$\rm_{LSR}$ $\approx$ 5 to 70 km~s$^{-1}$. The decomposition of the 625 GHz line into three hyperfine components, the main $F=5/2-3/2$ and the two satellite $F=1/2-3/2$ and $F=3/2-3/2$ hyperfine components, is also plotted in Figure \ref{fig4}. We add the three hyperfine components, shifting (in velocity) the satellite components accordingly to the main hyperfine component velocity shift, and obtain a hyperfine deconvolved spectrum, equivalent to a splitting-free spectrum, where each HFS component contributes to the optical depth. In order to estimate the line width and the center velocity of each individual velocity component, we fit the spectrum with a set of Gaussian curves (see Figure \ref{fig5}). The five Gaussian components used in our fit were set to give a better correspondence to other molecular transitions observed towards the same background source, such as H$_2$Cl$^+$ \citep{Neu12} and CH \citep{Ger10}. 
For comparison, we also plot in Figure \ref{fig5} the H$_2$$^{35}$Cl$^+$ and H$_2$$^{37}$Cl$^+$ HFS deconvolved absorption profiles from \cite{Neu12} and the H I absorption spectrum \citep{Fis03}. The H$_2$$^{35}$Cl$^+$ and H$_2$$^{37}$Cl$^+$ profiles are in good agreement with that of H$^{35}$Cl for most velocity intervals, as expected from precursor species of HCl. 
  

\subsection{Column densities}

We calculate the hydrogen chloride column densities for the velocities intervals associated with the different Gaussian components. First, we derive optical depths of the HCl lines ($\tau$ = -ln[2T$_{L}$/T$_{C}$-1]), assuming that the foreground absorption completely covers the continuum source and that all HCl molecules are in the ground state. We derive the H$^{35}$Cl column densities for each LSR velocity range using equation 3 of \cite{Neu10}, where the absorption optical depth for the H$^{35}$Cl $J = 0 -1$ transition, integrated over velocity, is given by
\begin{equation}
	\int \tau d \upsilon = \frac{A_{ul}g_u \lambda^3}{8\pi g_l}N({\rm H^{35}Cl})\Rightarrow  N({\rm H^{35}Cl})= 6.5211~\times~10^{12}\int \tau d \upsilon \left[cm^{-2}\right] \nonumber
\end{equation}
where $g_u$~=~3 and $g_l$~=~1 are the degeneracies of the upper ($J=1$) and lower ($J=0$) state. Since we use the deconvolved spectrum without hyperfine splitting, $\lambda$~=~478.96~$\mu$m is the transition wavelength, and $A$$\rm_{ul}$~=~1.17~$\times$~10$^{-3}$ s$^{-1}$ is the spontaneous radiative decay rate. 
Table 2 shows the HCl column densities and the abundances with respect to the total hydrogen column density. We calculate the $^{35}$Cl/$^{37}$Cl ratio in the velocity interval free from emission line contamination (shaded interval in Figure \ref{fig2}), which corresponds to velocities from 38 to 40.5 km s$^{-1}$. The $^{35}$Cl/$^{37}$Cl ratio obtained is equal to 2.9, in good agreement with the solar value of 3.1 \citep{And89}. Thus, the HCl abundances are calculated with respect to the total molecular and neutral hydrogen assuming a $^{35}$Cl/$^{37}$Cl of 3.1 in all velocity intervals. The H I column densities in the foreground gas are obtained from \cite{Fis03}, while the molecular hydrogen is obtained from CH \citep{Ger10}, assuming a linear correspondence between CH and H$_2$ with a scaling factor of 3.5 $\times$ 10$^{-8}$ \citep{She08}. The average HCl fractional abundance with respect to the total hydrogen is 6.1~$\pm$ 0.6~$\times$~10$^{-10}$. Using a chlorine abundance in diffuse clouds, $N$(Cl)/$N$$\rm_H$, of 1.03~$\times$~10$^{-7}$ measured by \cite{Moo12}  towards nearby (a few hundred pc) stars, we obtain that the average fraction of gas-phase Cl in HCl is $\sim$~0.6~\%. 

The uncertainties in the column densities are the random noise and the systematic errors introduced by the calibration uncertainties and the robustness of the HFS decomposition procedure. The 1-$\sigma$ uncertainty in the column density is dominated by the error in the linear baseline fitting. The total HIFI calibration uncertainties for band 1b are $\leq$~9.3~\%,  which includes the contribution from the sideband gain ratio, beam efficiency,  pointing, hot and cold coupling, where all errors are added in quadrature. Detailed information about the HIFI calibration can be found on \cite{Roe12}. The error introduced by the robustness of the deconvolution procedure is $\sim$ 14~\% \citep{God12}. Adding in quadrature all the errors, the total uncertainty in the column densities is about 19 \%.

 \section{Chemical model}
 
We use the Meudon PDR code \citep{LeP06} to compute the fractional abundances of the various chlorine bearing species, i.e. the abundance of each species normalized relative to the chlorine elemental abundance, as a function of the total extinction of the cloud, $A$$\rm_{v, tot}$. The Meudon PDR code is a one- or two-sided model, where the molecular cloud is modeled as a stationary plane-parallel slab of gas and dust, exposed to incident radiation field from a bright star or the Interstellar Standard Radiation Field (ISRF). As input parameters, we use the best fit model from \cite{Neu12}, which corresponds to an initial density of $n\rm_H$ = 316 cm$^{-3}$, a radiation field intensity from both sides of 10 and a primary ionization rate of 1.810~$\times$~10$^{-16}$~s$^{-1}$. Using the standard Galactic gas-to-dust ratio $N$$\rm_H$/$A$$\rm_v$~=~1.93~$\times$~10$^{21}$ mag$^{-1}$~cm$^{-2}$ \citep{Whi03}, $A$$\rm_{v, tot}$ probed is $\sim$~3~--~5. 
The obtained abundances normalized relative to the relevant gas-phase elemental abundance are shown in Figure 6-\emph{left}, the results indicate that the HCl$^+$ and H$_2$Cl$^+$ account jointly at best for $\sim$ 2~\% of the elemental chlorine in the gas phase while HCl accounts for $\sim$ 0.1~\%. The modeled HCl fractional abundance under predicts the results obtained from the observations by an average value of about 6.   

The DR studies of H$_2$Cl$^+$ and the corresponding product branching ratios are not yet available from laboratory experiments. In their models, NW09 assumed a branching ratio of 90~\% for Cl and 10~\% for HCl, motivated by the low HCl abundance observed towards the diffuse cloud $\zeta$~Oph. We have thus investigated the effect of the branching ratio of H$_2$Cl$^+$ DR for the physical conditions described above, based on the correlation obtained between the branching ratio of triatomic dihydride ions for 3-body dissociation and the corresponding energy release obtained by \cite{Rou11}. From that correlation, \cite{Rou11} obtain a branching ratio of 56~\% for the 3-body dissociation of H$_2$Cl$^+$. We estimate the fractional abundances corresponding to the new branching ratio and plot them as dashed lines in Figure 6-\emph{right}, in comparison with those obtain with previous assumption (90 \% for Cl) in full lines, as a function of $A$$\rm_{v_{tot}}$. The results show that differences occur only for the minor species, i.e. HCl, HCl$^+$ and H$_2$Cl$^+$ while the reservoirs Cl and Cl$^+$ are not affected. The increase in the HCl column density is about a factor of $\sim$~5, in closer agreement with our observational results. However, the increase for HCl$^+$ and H$_2$Cl$^+$ is much smaller, and then this result cannot explain all the discrepancies. 

\section{Discussion}

We present the detection of H$^{35}$Cl and H$^{37}$Cl $J = 0 - 1$ transition in absorption on the line of sight towards the bright submillimeter source W31C. The obtained HCl column density ($\sim$~2~$\times$~10$^{13}$ cm$^{-2}$), is comparable with that obtained towards the line of sight to Sgr B2 (S) \citep{Lis10}. Given the HCl$^+$ and H$_2$Cl$^+$ column densities from \cite{deL12} and \cite{Neu12}, the HCl$^+$/H$_2$Cl$^+$ and HCl/H$_2$Cl$^+$ column density ratios, of $\sim$~1 and $\sim$~0.3 respectively, are within the range predicted by the models of diffuse clouds. However, the averaged HCl abundance with respect to the total hydrogen content of 6.1~$\pm$ 0.6~$\times$~10$^{-10}$ suggests a HCl fractional abundance with respect to the chlorine elemental abundance ($N$(Cl)/$N$$\rm_H$~=~1.03~$\times$~10$^{-7}$)  a factor of about 6 larger than those predicted by chemical models of diffuse clouds. Similar discrepancies between models and observations were also found for other Cl-bearing gas-phase species, HCl$^{+}$ and H$_2$Cl$^+$ towards W31C, Sgr A (+50 km s$^{-1}$ molecular cloud) and W49N \citep{Lis10,deL12, Neu12}. 

\cite{Lis10} suggested a geometrical explanation for the discrepancy between the observed column densities and the models, where enhancement in the absorbing column density could result from multiple PDRs present along the line of sight. This scenario could explain higher column densities, but the HCl/Cl ratio will remain constant and hence this solution is not likely to reproduce the W31C line of sight results. 

\cite{Neu12} explored several possible explanations for the deficiencies in the model predictions, e.g. the rate coefficient assumed for dissociative recombination (DR) of HCl$^{+}$ and uncertainties in the assumed dipole moment of H$_2$Cl$^+$, none of which proved satisfactory. We have investigated the effect of varying the branching ratio of H$_2$Cl$^+$ DR. By taking a branching ratio of 56~\% for the chlorine (instead of 90~\% assumed in NW09 models), the predicted fractional abundance of HCl with respect to the gas-phase chlorine abundance increases by about a factor of $\sim$~5. However, laboratory results are needed to confirm the branching ratio and validate this argument, along with the study of other unknowns in the chlorine chemistry, such as the DR rates, and the dipole moment of H$_2$Cl$^+$.

\acknowledgments

HIFI has been designed and built by a consortium of institutes and university departments from across Europe, Canada and the United States (NASA) under the leadership of SRON, Netherlands Institute for Space Research, Groningen, The Netherlands, and with major contributions from Germany, France and the US. Consortium members are: Canada: CSA, U. Waterloo; France: CESR, LAB, LERMA, IRAM; Germany: KOSMA, MPIfR, MPS; Ireland: NUI Maynooth; Italy: ASI, IFSI-INAF, Osservatorio Astrofisico di Arcetri-INAF; Netherlands: SRON, TUD; Poland: CAMK, CBK; Spain: Observatorio Astronomico Nacional (IGN), Centro de Astrobiologia; Sweden: Chalmers University of Technology - MC2, RSS \& GARD, Onsala Space Observatory, Swedish National Space Board, Stockholm University - Stockholm Observatory; Switzerland: ETH Zurich, FHNW; USA: CalTech, JPL, NHSC. Support for this work was provided by the Centre National de Recherche Spatiale (CNES), by the SCHISM project (grant ANR-09-BLAN-0231-01), by NASA through an award issued by JPL/Caltech, and by the Spanish MICINN (grants AYA2009-07304 and CSD2009-00038). This research has been supported in part by the NSF, award AST-0540882 to the CSO. 

{\it Facilities:} \facility{Herschel/HIFI}.

\begin{center}
\begin{deluxetable}{cccccc}
\tabletypesize{\scriptsize}
\tablecaption{LTE model parameters.}
\tablewidth{270pt}
\tablehead{
Species&Source size&$T\rm_{ex}$&$\upsilon$$\rm_{LSR}$&FWHM& $N$\\
&(")&(K)&(km s$^{-1}$)& (km s$^{-1}$) & (cm$^{-2}$)
}
\startdata
 CH$_3$OCH$_3$&33&60&-2.5&5.5&2~$\times$~10$^{14}$\\[2pt]
 CH$_3$CN&4&150&-1&5.5&7.4~$\times$~10$^{14}$\\[2pt]
  SO$_2$&4&150&-1&5&6.5~$\times$~10$^{15}$\\[2pt]
\enddata
\end{deluxetable}
\end{center}

\clearpage

\begin{center}
\begin{deluxetable}{cccccc}
\tabletypesize{\scriptsize}
\tablecaption{HCl column densities and abundances on the line of sight towards W31C.}
\tablewidth{350pt}
\tablehead{
$v\rm_{LSR}$&$N$(H I)\tablenotemark{a}&$N$(H$_2$)\tablenotemark{b}&$N$(H$^{35}$Cl)&$N$(HCl)\tablenotemark{c}&$N$(HCl)/$N$$\rm_H$\tablenotemark{d}
\\
km s$^{-1}$&  $\times$~10$^{21}$~cm$^{-2}$&$\times$~10$^{21}$~cm$^{-2}$&$\times$~10$^{12}$~cm$^{-2}$&$\times$~10$^{12}$~cm$^{-2}$&$\times$~10$^{-10}$
}
\startdata
		  $\left[ 15,19 \right]$  & 1.3 $\pm$ 0.35 &2.9 $\pm$ 0.3&2.8 $\pm$ 0.51& 3.7 $\pm$ 0.67 &5.2 $\pm$ 1.1\\[2pt]
		  $\left[ 19,24 \right]$  &1.6 $\pm$ 0.56&2.7 $\pm$ 0.3&3.8 $\pm$ 0.69& 5.0 $\pm$ 0.91&7.1 $\pm$ 1.5\\[2pt]
		  $\left[ 25,29 \right]$  &1.6 $\pm$ 0.40&3.6 $\pm$ 0.3&1.3 $\pm$ 0.28& 1.7 $\pm$ 0.37&1.9 $\pm$ 0.5 \\[2pt]
		  $\left[ 30,36 \right]$  & 2.1 $\pm$ 0.78&3.5 $\pm$ 0.3& 4.0 $\pm$ 0.74& 5.3 $\pm$ 0.98&5.9 $\pm$ 1.2\\[2pt]
		  $\left[ 36,40 \right]$  &1.7 $\pm$ 0.20&3.0 $\pm$ 0.3& 6.0 $\pm$ 1.02& 8.0 $\pm$ 1.3&10.3 $\pm$ 1.9\\[2pt]
		  $\left[ 15,40 \right]$  &8.4 $\pm$ 1.12&15.6 $\pm$ 0.7& 17.8 $\pm$ 1.2& 23.6 $\pm$ 2.1&  6.0 $\pm$ 0.6\\[2pt]
\enddata
\tablenotetext{a}{The H I column densities are obtained from \cite{Fis03} absorption data with $N$(H~I)~=~1.84$\times$10$^{20}$(T$\rm_{spin}$/100K)$\int$~$\tau_{\rm HI}$d$\upsilon$ [cm$^{-2}$/km~s$^{-1}$], assuming a T$\rm_{spin}$~$\sim$~100~K}.
\tablenotetext{b}{The H$_2$ column densities are derived from CH observations \citep{Ger10}, assuming $N$(CH) = 3.5$\times$10$^{-8}$ $N$(H$_2$) \citep{She08}.}
\tablenotetext{c}{Assuming a solar $^{35}$Cl/$^{37}$Cl ratio of 3.1}
\tablenotetext{d}{$N$$\rm_H$=2$N$(H$_2$)+$N$(H~I)}
\end{deluxetable}
\end{center}
\clearpage

\clearpage

\clearpage

\clearpage

\clearpage

\clearpage

\begin{center}
\begin{tabular}{cc}
\includegraphics[angle=0,scale=.30]{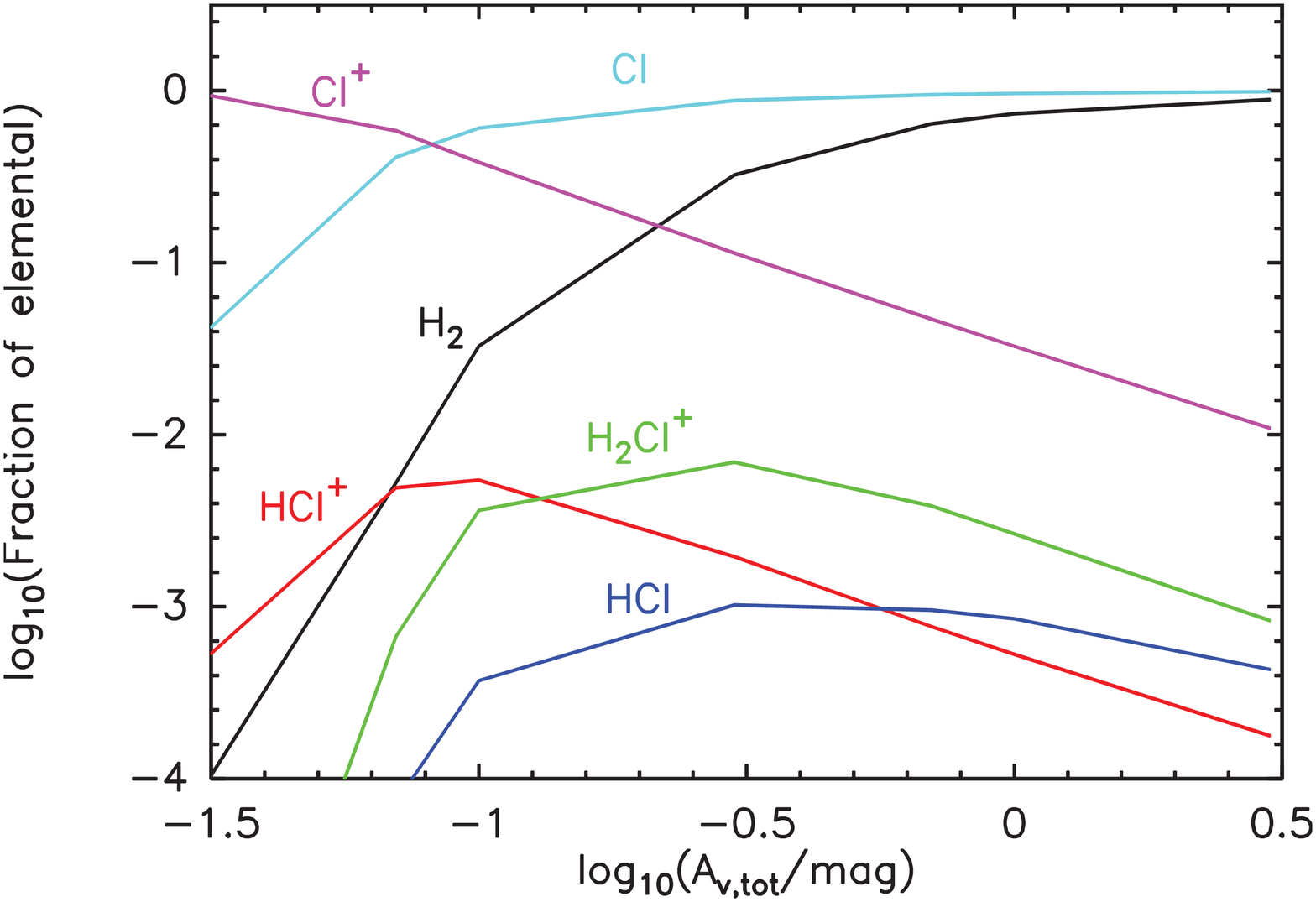}&
\includegraphics[angle=0,scale=.30]{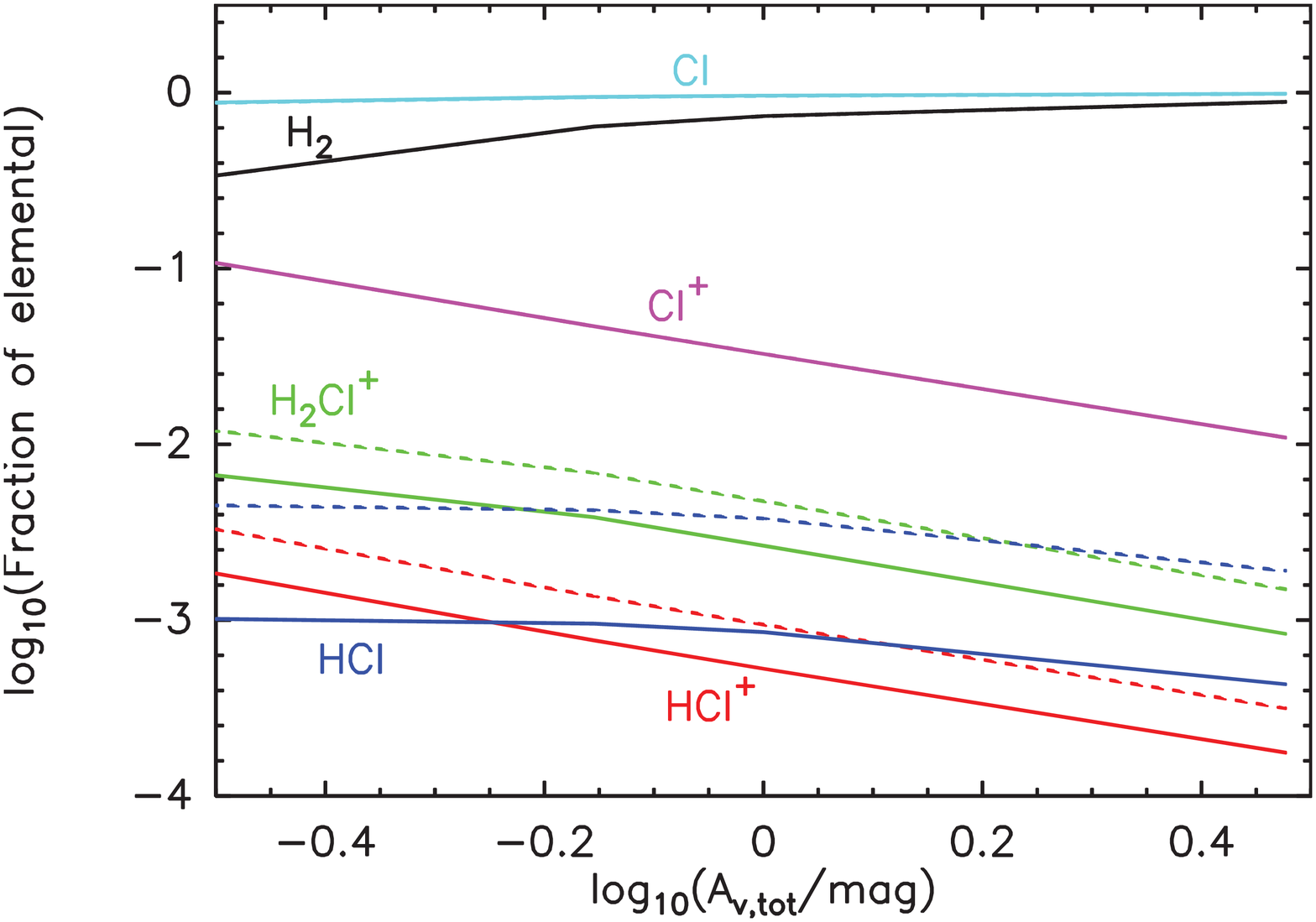}\\
\end{tabular}

\end{center} {Fig. 6.---  Left, abundance of the Cl-bearing molecules, chlorine atom or ion (X), normalized relative to the gas-phase chlorine abundance, $\overline{f}$(X)~=~$N$(X)/$N$(Cl), obtained with the Meudon PDR code, Cl$^+$ (magenta), Cl (cyan), HCl (blue), HCl$^{+}$ (red), and H$_2$Cl$^+$ (green) together with the H$_2$ fractions (black), $\overline{f}$(X)~=~2$N$(H$_2$)/(2$N$(H$_2$)+$N$(H I)), averaged along the line of sight, as a function of total visual extinction $A\rm_{v,tot}$. The initial density is $n_{\rm H}$~=~316~cm$^{-3}$, and the radiation field intensity is 10. Right, comparison of the predicted fraction of gas-phase chlorine present in HCl, HCl$^+$, H$_2$Cl$^+$, Cl and Cl$^+$ assuming a branching ratio of 90 \% for chlorine and 10 \% for hydrogen chloride (full lines) and the predicted fractions for a reduced branching ratio of 56 \% for chlorine (dashed line). \label{fig6}}

\end{document}